# 3D Structural Analysis of the Optic Nerve Head to Robustly Discriminate Between Papilledema and Optic Disc Drusen

**Michaël J.A. Girard**[1,2,3], **Satish K. Panda**[1], **Tin A. Tun**[2,4], **Elisabeth A. Wibroe**[5], **Raymond P. Najjar**[2,4,6], **Tin Aung** [2,4,6], **Alexandre H. Thiéry**[7], **Steffen Hamann**[5], **Clare Fraser**[8], and **Dan Milea**[2,4]

1. Ophthalmic Engineering & Innovation Laboratory, Singapore Eye Research Institute, Singapore National Eye Centre, Singapore
2. Duke-NUS Graduate Medical School, Singapore
3. Institute for Molecular and Clinical Ophthalmology, Basel, Switzerland
4. Singapore Eye Research Institute, Singapore National Eye Centre, Singapore
5. Department of Ophthalmology, Rigshospitalet, University of Copenhagen, Denmark
6. Yong Loo Lin School of Medicine, National University of Singapore, Singapore
7. Department of Statistics and Applied Probability, National University of Singapore, Singapore
8. Save Sight Institute, Faculty of Health and Medicine, The University of Sydney, Sydney, New South Wales, Australia

# Abstract

**Purpose**: **(1)** To develop a deep learning algorithm to identify major tissue structures of the optic nerve head (ONH) in 3D optical coherence tomography (OCT) scans; **(2)** to exploit such information to robustly differentiate among healthy, optic disc drusen (ODD), and papilledema ONHs.

**Methods:** This was a cross-sectional comparative study with confirmed ODD (105 eyes), papilledema due to high intracranial pressure (51 eyes), and healthy controls (100 eyes). 3D scans of the ONHs were acquired using OCT, then processed to improve deep-tissue visibility. In a $1^{st}$ step, a deep learning algorithm was developed using 984 B-scans (from 130 eyes) in order to identify: major neural/connective tissues, and ODD regions. The performance of our algorithm was assessed using the Dice coefficient (DC). In a $2^{nd}$ step, a classification algorithm (random forest) was designed using 150 OCT volumes to perform 3-class classifications (1: ODD, 2: papilledema, 3: healthy) strictly from their drusen and prelamina swelling scores (derived from the segmentations). To assess performance, we reported the area under the receiver operating characteristic curves (AUCs) for each class.

**Results.** Our segmentation algorithm was able to isolate neural and connective tissues, and ODD regions whenever present. This was confirmed by an averaged DC of 0.93±0.03 on the test set, corresponding to good performance. Classification was achieved with high AUCs, i.e. 0.99±0.01 for the detection of ODD, 0.99±0.01 for the detection of papilledema, and 0.98±0.02 for the detection of healthy ONHs.

**Conclusions:** Our AI approach allows us to accurately discriminate ODD from papilledema, using a single OCT scan. Our classification performance was excellent, with the caveat that validation in a much larger population is warranted. Our approach may have the potential to establish OCT as the mainstay of diagnostic imaging in neuro-ophthalmology.

# Introduction

The optic nerve head (ONH), also commonly called the optic disc, is a fragile structure located at the junction between the optic nerve and the eye globe,[1], providing information about the brain's health.[2] The ONH contains retinal neuronal axons connecting the eye to the brain. In patients with neurological conditions causing high intracranial pressure (tumors, vascular abnormalities or pseudotumor cerebri), axoplasmic stasis will cause these axons to swell, causing papilledema. Papilledema can be observed clinically using ophthalmoscopy, but its distinction (especially at early stages) from other ONH lesions mimicking papilledema (i.e. pseudopapilledema) can be difficult in clinical practice. In recent years, imaging of papilledema has benefited from the use of optical coherence tomography (OCT) – a 3D imaging modality that allows fast, high-resolution and non-invasive visualization of the ONH. OCT allows early detection of disc swelling and can also be used in the diagnosis of the most common cause of pseudo-papilledema, optic disc drusen (ODD; **Figure 1**).[3] This distinction is critical in everyday practice, for early and prompt detection of potentially life-threatening brain conditions, or on the other hand, to avoid unnecessary, invasive, expensive investigations in patients with ODD.[4]

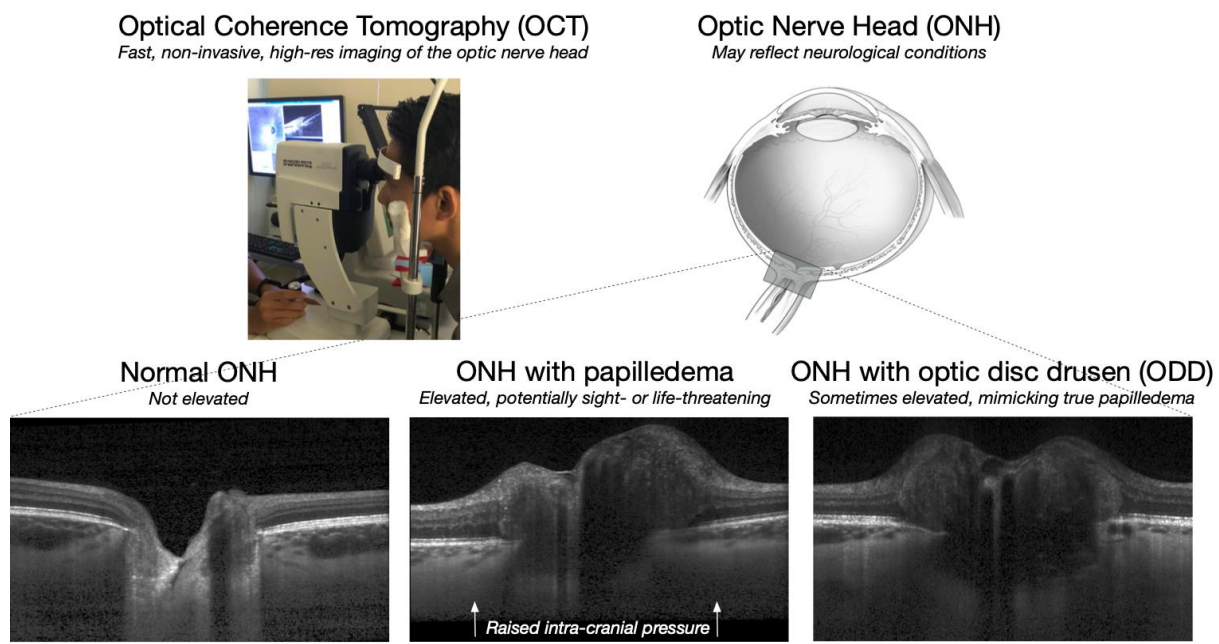

**Figure 1.** The optic nerve head (ONH), located at the back of the eye, can reflect neurological conditions such as papilledema. This condition is potentially sight- or life-threatening but hard to differentiate from a less dangerous condition known as optic disc drusen (ODD).

To date, there is no clear medical imaging protocol to differentiate ODD from papilledema. While OCT is gaining ground due to its high resolution and availability,[4-6] many clinicians continue to rely on earlier technologies, including: B-scan ultrasonography, fundus autofluorescence imaging, or fluorescein angiography.[7-11] If papilledema is suspected, referral to a neuro-ophthalmologist or neurologist is critical, for further evaluation of its cause using expensive neuroimaging and invasive measurements of the opening pressure of the cerebrospinal fluid via lumbar puncture. Such investigations are not warranted in ODD patients, whose disc appearance can nevertheless mimic papilledema and thereby create a clinical dilemma. In short, identification of ODD remains a clinical challenge, which can benefit from new technologies for improved discrimination from papilledema, that could be sight- or life-threatening.

While numerous recent studies have confirmed the clinical utility of OCT to differentiate ODD from papilledema,[5, 6, 12-17] none have yet identified novel and robust OCT-derived biomarkers (other than retinal layer thicknesses [18]), or exploited the complex structural signature of a given ONH to automatically dissociate these 2 conditions with high accuracy.

In this study, we aimed to develop a deep learning algorithm to automatically and simultaneously identify all major tissue structures of the ONH (including neural & connective tissues, and ODD regions). This information was then exploited in a separate population by a subsequent AI algorithm to differentiate among healthy, ODD, and papilledema ONHs. We ultimately hope to provide a clinical tool to assist both ophthalmologists and neurologists in their practice.

## Methods

### Patient Recruitment

A total of 241 subjects (for a total of 256 ONHs, i.e. 105 ODD, 51 papilledema, and 100 healthy ONHs) were retrospectively included in this study at three different sites: **(1)** the Singapore National Eye Centre (SNEC, Singapore), **(2)** Rigshospitalet, University of Copenhagen (Glostrup, Denmark) and **(3)** the Save Sight Institute, Faculty of Health and Medicine, University of Sydney (Sydney, Australia). The cohort at SNEC was composed of 5 ODD, 1 papilledema, and 100 healthy discs. The cohort from Rigshospitalet included 90 ODD, and 50 papilledema; that from the Save Sight Institute included 10 ODD. All subjects gave written informed consent. The study adhered to the tenets of the Declaration of Helsinki and was approved by the institutional review board of the respective institutions. A summary of the patient populations and their demographics is shown in **Table 1**.

|  | Country | Age | Sex (%Male) | ODD Scans | Papilledema Scans | Healthy Scans | Total |
| --- | --- | --- | --- | --- | --- | --- | --- |
| Cohort 1 | Denmark | 35±17 | 21% | 90 | 50 | 0 | 140 |
| Cohort 2 | Australia | 45±26 | 60% | 10 | 0 | 0 | 10 |
| Cohort 3 | Singapore | 59±7 | 49% | 5 | 1 | 100 | 106 |
|  | Total |  |  | 105 | 51 | 100 | 256 |

**Table 1.** Summary of patient populations.

ODD was defined as hyporeflective structures on OCT of the ONH according to the guidelines provided by the Optic Disc Drusen Studies (ODDS) Consortium.[4] None of the patients with ODD had other ONH pathologies including concurrent papilledema.

Papilledema was defined as bilateral ONH swelling associated with high intracranial pressure, which was either primary (pseudotumor cerebri) or secondary (due to space occupying lesions, brain tumors, venous sinus thrombosis, etc). Raised intracranial pressure was confirmed either by the presence of a space occupying lesion on neuroimaging and/or by elevated opening pressure on lumbar puncture in standard conditions.

**Optical Coherence Tomography Imaging**

OCT imaging was performed on seated subjects in a dark room, if necessary after pupil dilation with tropicamide 1% solution. All 3 imaging sites (Denmark, Australia, Singapore), used the same spectral-domain OCT device (Spectralis, Heidelberg Engineering, Germany) with enhanced depth imaging (EDI) enabled. Some variations were present in our acquisition parameters. While all OCT volumes (all horizontal raster scans) covered the entire ONH, the number of B-scans (slices) varied between 25 and 193, and the number of A-scans per B-scan between 384 and 1,024. The number of pixels per A-scan was always fixed at 496. In terms of resolution, the distance between B-scans varied between 27.3 and 246.0 μm, the lateral resolution between 5.5 and 13.1 μm, and the axial resolution was always fixed at 3.9 μm. Signal averaging was used for all OCT scans and varied between 10 and 30.

**Correction of Light Attenuation Using Adaptive Compensation**

In order to remove the deleterious effects of light attenuation from OCT images, all B-scans were post-processed using adaptive compensation.[19, 20] For OCT images of the ONH, adaptive compensation has been shown to remove blood vessel shadows, improve tissue contrast, and increase deep-tissue visibility, such as that of the lamina cribrosa (LC). For all B-scans, we used a threshold exponent of 12 (to limit noise over-amplification at high depth), and a contrast exponent of 2 (to improve overall image contrast). This step was also found critical to visualize and manually segment the LC in swollen discs, and to better identify ODD regions and their conglomerates.

**Manual Segmentation of OCT Images as Required for AI Training**

We performed manual segmentation of a subset of compensated OCT images in order to: **1)** train and test a segmentation algorithm to highlight neural, connective, and drusen tissues; and in a subsequent step to **2)** use such information to discriminate papilledema from ODD. Specifically, we manually segmented a total of 984 OCT images (B-Scans), including 167 from ONHs exhibiting ODD (taken from 50 eyes), 108 from papilledema ONHs (from 30 eyes), and 709 from healthy ONHs (from 50 eyes). Note that this dataset (a sub-population of our recruited patients) was referred to as the segmentation dataset. Briefly, each compensated OCT image was manually segmented by one expert observer (MJAG) using Amira (version 5.4, FEI, Hillsboro, OR) to identify the following tissue groups: **(1)** the retinal nerve fiber layer (RNFL) and the prelamina (**Figure 2**); **(2)** the ganglion cell layer and the inner plexiform layer (GCL+IPL); **(3)** the retinal pigment epithelium (RPE) with Bruch's membrane (BM); **(4)** all other retinal layers; **(5)** the choroid; **(6)** the peripapillary sclera including the scleral flange; **(7)** the lamina cribrosa; and whenever present **(8)** ODD regions and conglomerates, including their hypo-reflective core and hyper-reflective margins.

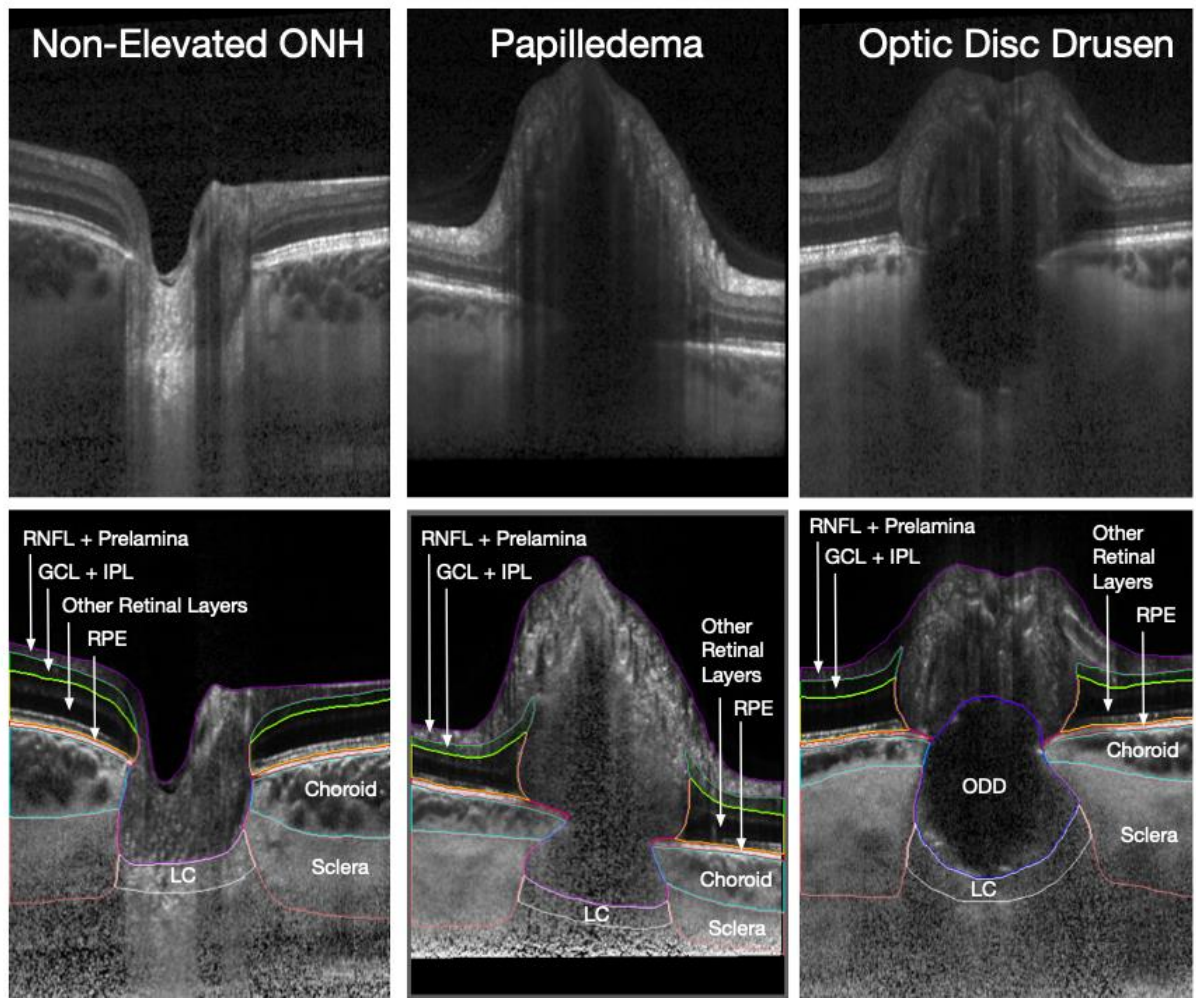

**Figure 2.** To train our AI algorithm, manual segmentation was performed on OCT images from healthy, ODD, and papilledema subjects. The following tissues (or tissue groups) were identified: (1) RNFL + prelamina, (2) GCL+IPL, (3) other retinal layers, (4) RPE, (5) choroid, (6) sclera, (7) lamina cribrosa, and (8) ODD regions and conglomerates whenever present. Tissue boundaries are shown on enhanced (compensated) images in the 2$^{nd}$ row, while baseline images (non-processed) are shown in the 1$^{st}$ row.

When multiple conglomerates were co-adjacent, they were fused in the manual segmentation for simplicity, rather than attempting to isolate them as separate islands. Note that we did not attempt to segment the peripapillary hyper-reflective ovoid mass-like structures (also known as PHOMS), as their boundaries were found to be subjective in both ODD and papilledema eyes. Note also that in most cases (especially swollen discs), we could not achieve full-thickness segmentation of the peripapillary sclera and of the LC due to limited visibility at high depth, even when compensation was used.[21] Therefore, we only segmented the OCT-visible portions of the sclera and LC as per the observed compensated signal, and no efforts were made to capture their accurate thickness. The manual segmentation assigned a label (between 1 and 8; background: 0) to each pixel of each OCT image to indicate the tissue class.

## Automated Segmentation of Neural, Connective, and Drusen Tissues using Deep Learning

To automatically segment all tissue layers of the ONH, we used a Unet++ model with ResNet-34 as backbone, implemented in Pytorch.[22] Unet++ is a fully convolutional neural network for semantic segmentation consisting of an encoder, a decoder, and a series of "nested, dense skip pathways", with a performance superior to that of U-Net.[23] The encoder extracted features from the input images that were used by the decoder to generate segmentation masks, whereas the skip connections were used to avoid the vanishing gradient problem and to facilitate the backpropagation process. We used the ResNet-34 network with pre-trained weights (on ImageNet) for the encoder. Note that ResNet-34 is one of the state-of-the-art networks that exploits the power of residual connections.[24] The number of output channels of the decoder was reduced to nine (i.e. the number of tissue classes including the background) with a SoftMax activation applied to the final convolution layer. We used the Jaccard index (mean for all tissues) to represent the loss function. To avoid overfitting during the training process, we performed extensive data augmentation using the python library Albumentation.[25] Image transformations such as horizontal flipping, random rotation and translation, additive Gaussian noise, random brightness, contrast, blur, and saturations were used as data augmentation in order to enrich the training set. Such transformations were found to considerably improve segmentation performance.

Images from the segmentation dataset were split into training (70%), validation (15%), and test (15%) sets, respectively. The split was performed in such a way that duplicate images or images from the same subject did not exist in different sets. We also ensured that ODD, papilledema, and healthy images were present in all sets according to the 70%/15%/15% split. For instance, in the training set, images from 35 ODD, 21 papilledema, and 50 healthy ONHs were used. All B-scan images were resized to 256x256 pixels to normalize different scan types. Our network was then trained on a Nvidia 1080Ti GPU card until optimum performance was reached in the validation set in about 100 epochs (computational time: ~1 hour). To evaluate the accuracy of our segmentation network, we calculated the Dice coefficient by comparing the network predicted labels with the corresponding manually segmented images from the test set. A Dice coefficient of one indicated a perfect match between the network prediction and the manual segmentation, while 0 indicated no overlap. For the Dice coefficient, we reported the mean ± standard deviation through a 5-fold cross-validation process.

## Classification of each OCT Volume as Healthy, Papilledema, or ODD

In this work, we aimed to perform multiclass classifications to identify whether a given ONH (described as a 3D volume with OCT) would be classified as healthy, papilledema, or ODD. For such classifications, we opted to use a simple solution in which we exploited the output of the deep-learning segmentations. Once such segmentations were obtained, we computed the Drusen Score and the Prelamina Swelling score for each OCT volume before deciding on classification. Both scores were computed as follows:

**Drusen Score.** The Drusen Score was simply defined as the total volume of ODD regions and conglomerates (expressed in $mm^3$). It was meant to easily help identify ODD eyes. For each given ONH, we first computed the total volume of ODD regions and conglomerates by computing the total number of pixels belonging to the ODD class (class 8). As each OCT volume had a different resolution, we multiplied the total number of pixels by the corresponding pixel width, pixel height, and the distance between B-scans for any given OCT volume. A Drusen score of 1 $mm^3$ (or 1 mm x 1 mm x 1 mm) would be considered large as could occur in the presence of an ODD region covering the entire LC and that is present both anteriorly and posteriorly to Bruch's membrane opening (BMO). Conversely, we would expect a Drusen score of 0 $mm^3$ for both healthy and papilledema ONHs.

**Prelamina Swelling Score.** The Prelamina Swelling Score was defined as the prelamina volume (including ODD regions; expressed in $mm^3$) contained within a cylinder with BMO as its circumference. Such a score was defined because a large score can easily help identity swollen discs as typically observed in papilledema. For each OCT volume, we simply identified all the prelamina (class 1) and ODD (class 8) pixels that did not cover the RPE (class 4) in the en-face view. The total number of identified pixels was then multiplied by the corresponding pixel width, pixel height, and the distance between B-scans for each OCT volume.

**Classification with Random Forest.** To avoid any bias, it was critical to address the problem of classification with a dataset that was independent from that used to train our segmentation network. Our dataset used for classification (referred to as the classification dataset – a subpopulation of our recruited subjects) was composed of 70 ODD, 30 papilledema, and 50 healthy OCT volumes. Using such data, a machine learning classifier, i.e., random forest (RF), was used to perform three-class classifications (class 1: ODD, class 2: papilledema, class 3: healthy) solely from the drusen and prelamina swelling scores computed on each OCT volume. The classification dataset was split into 50% training and 50% test sets, and then the RF classifier was trained in a supervised manner on the training set. Subsequently, the network was tested on the testing set. We reported the area under the receiver operating characteristic curves (AUCs) for each class using the one-vs-all approach, i.e., ODD vs Papilledema+Healthy, Papilledema vs ODD+Healthy, and Healthy vs ODD+Papilledema. Finally, five-fold cross-validation was performed to assess the performance of our network, and all AUCs were reported as mean ± standard deviation.

# Results

## ONH Segmentation: Qualitative and Quantitative Analysis

Compensated, manually-segmented, and AI-segmented images (all from the test set) for 6 selected ONH conditions (2 from Healthy, 2 from Papilledema, and 2 from ODD subjects) can be found in **Figure 3**. Overall, our segmentation algorithm was able to simultaneously isolate neural tissues (RNFL+prelamina, GCL+IPL, RPE, and all other retinal layers), connective tissues (choroid, peripapillary sclera, and LC), and ODD regions/conglomerates whenever present. AI-based segmentations were in good agreement with those performed manually, even in the presence of heavy noise (**Figure 3**, healthy subject, 2$^{nd}$ Column). This was confirmed by an averaged Dice coefficient of 0.93±0.03 on the test set, corresponding to very good segmentation performance.

Errors were sometimes present in the AI-based segmentations. To name of few: **(1)** it was possible to observe partial segmentations of the LC (even though fully present in the manual segmentations) or even its complete absence; **(2)** Drusen shapes did not always matched those in the manual segmentations (**Figure 3**, ODD subject, 1$^{st}$ Column); **(3)** ODD regions were observed in 4 healthy images of the test set, but typically as small, isolated islands of a few pixels each. However, it was important to note that, in the segmentation test set, ODD regions were always identified in the ODD images, and no ODD pixels were identified in the papilledema images.

When applied to full 3D OCT volumes of the ONH, our algorithms were able to accurately highlight all tissue structures as continuous 3D entities, including swollen pre-laminas typically observed in papilledema eyes, and the ball-like appearance of ODD regions in ODD eyes. This was assessed qualitatively, by comparing 3D surface-meshed segmentations with 3D renderings of the OCT B-scans. For instance, two segmented volumes (1 ODD and 1 papilledema) and their corresponding 3D OCT renderings are shown in **Figure 4** and **Figure 5**, respectively.

## Classification

In the population used for classification, the mean Drusen score was 0.66±0.55 mm$^3$ for ODD eyes, 0.004±0.015 mm$^3$ for papilledema eyes, and 0.002±0.005 mm$^3$ for healthy eyes. The mean prelamina swelling score was 1.98±0.63 mm$^3$ for ODD eyes, 3.43±1.49 mm$^3$ for papilledema eyes, and 1.23±0.26 mm$^3$ for healthy eyes. The drusen and prelamina swelling scores for all 150 eyes used for classification are shown in **Figure 6**; 3 clusters, 1 for each class, can be observed.

Our classification training and test sets had 75 ONHs each. For the one-vs-all comparison, AUCs of 0.99±0.01 were obtained for the detection of ODD (vs healthy+papilledema), 0.99±0.01 for the detection of papilledema (vs healthy+ODD), and 0.98±0.02 for the detection of healthy ONHs (vs ODD+papilledema). The average classification accuracy was found to be 93±0.03%.

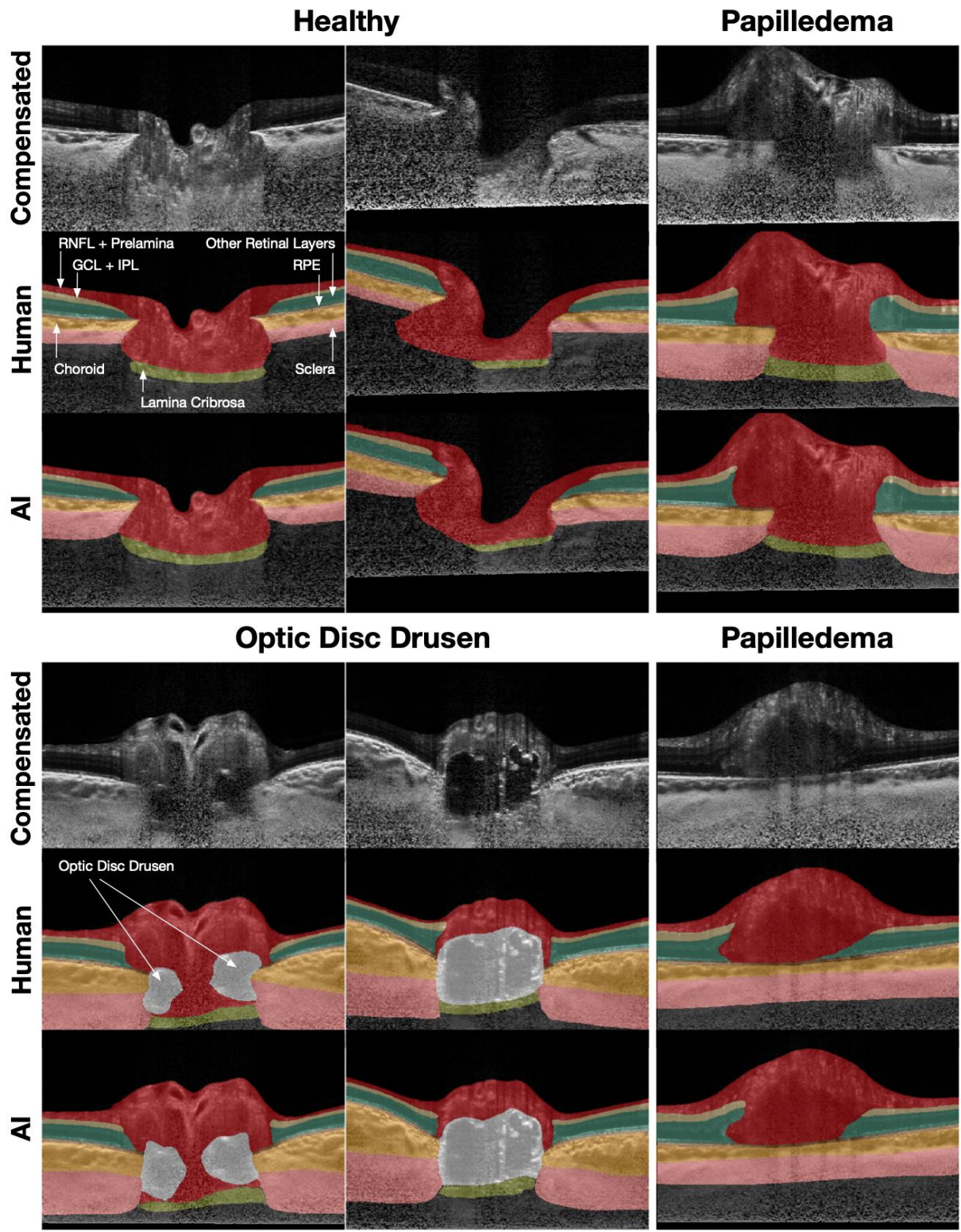

**Figure 3.** Compensated, manually-segmented, and AI-segmented for 6 selected ONHs (2 from healthy, 2 from papilledema, and 2 from ODD subjects). Overall, our segmentation algorithm was able to simultaneously isolate neural tissues (RNFL+prelamina, GCL+IPL, RPE, and all other retinal layers), connective tissues (choroid, peripapillary sclera, and LC), and ODD regions/conglomerates whenever present.

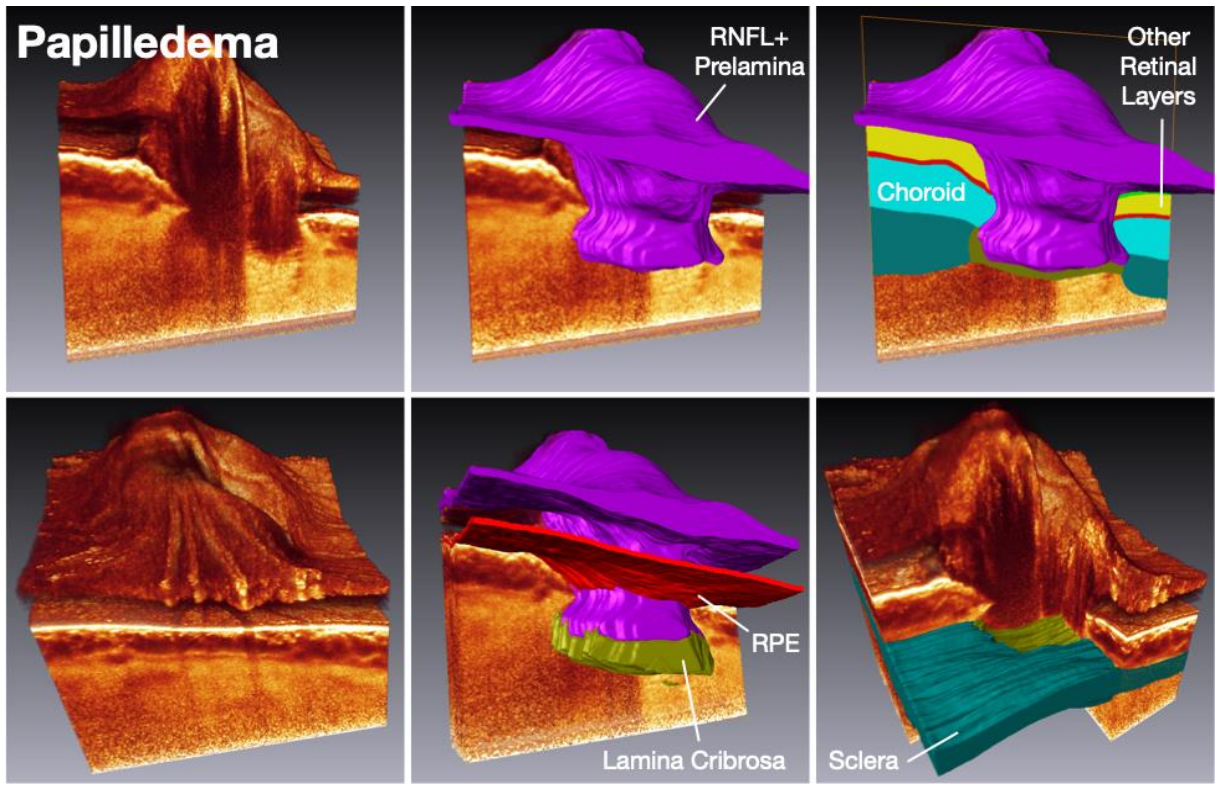

**Figure 4.** 3D rendering of one OCT volume and corresponding 3D surface-meshed segmentation (obtained with AI) for one papilledema ONH.

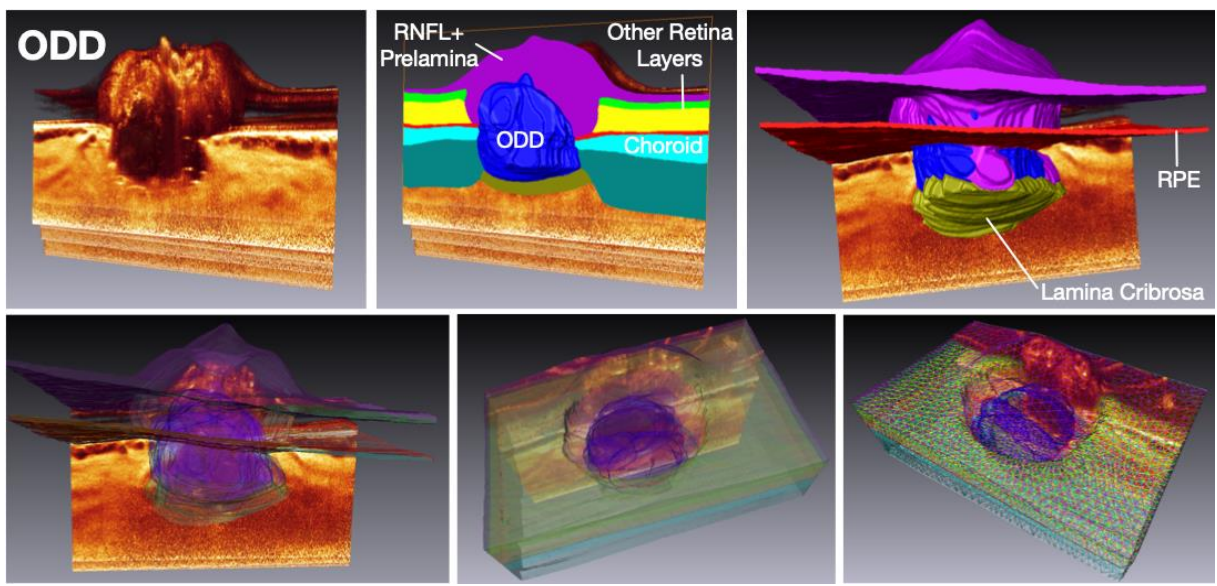

**Figure 5.** 3D rendering of one OCT volume and corresponding 3D surface-meshed segmentation (obtained with AI) for one ONH exhibiting ODD.

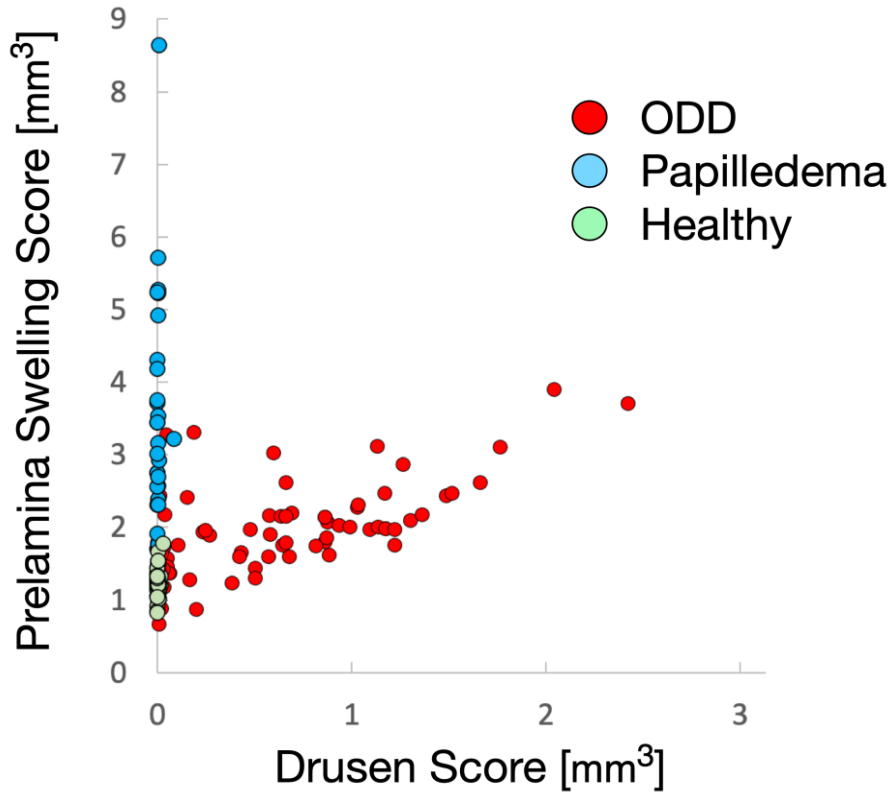

**Figure 6.** Drusen and prelamina swelling scores for all 150 eyes used for the classification problem.

## Discussion

In this study, we propose a relatively simple AI approach to discriminate ODD from papilledema strictly from a single OCT scan of the ONH. The assessment was performed in 2 steps. In the first, we identified all major neural and connective tissue structures of the ONH (and ODD regions whenever present) using deep learning. In the second, we evaluated the Drusen score and the prelamina swelling score (from the segmentations) to classify a given ONH as either healthy, ODD, or papilledema. Our classification performance was excellent, with the caveat that validation in a much larger population is warranted for clinical acceptance. Nevertheless, our approach may have the potential to establish OCT as the mainstay of diagnostic imaging for ODD and papilledema.

In this study, we obtained an averaged Dice coefficient of 0.93±0.03, which is comparable to past ONH segmentation performance obtained with various networks: 0.93±0.02 with ONH-Net,[26] 0.91±0.04 with DRUNET,[27] and 0.90±0.04 with U-Net.[28] The only difference being that ours added an additional class for ODD regions and conglomerates, which did not appear to decrease overall performance. In addition, and for the purpose of ODD segmentation, it is important to note that our network performed considerably better when 9 tissue groups were simultaneously identified in the OCT images, as opposed to 2 (i.e., ODD vs non-ODD regions). For this latter approach, a lower Dice coefficient of 0.86 was obtained.[29] Our work strongly suggests that correctly identifying all tissue groups, even those that are not primarily involved in major optic neuropathies, would in turn improve the detection of ODD regions. We could envision that adding additional classes, such as e.g. one for the central retinal vessel trunk (and its branches),[30] and one for PHOMS,[31] could further improve performance, as such tissues are strongly intertwined with ODD regions.

While our segmentation performance could be considered very good, it should be noted that the Drusen score was found to be non-zero in non-ODD eyes, and specifically: 0.004±0.015 mm$^3$ in papilledema eyes, and 0.002±0.005 mm$^3$ in healthy eyes. This suggests that ODD regions and conglomerates were identified in some non-ODD eyes, but in small quantities (as can be seen in Figure 6). To overcome this issue, one could consider: **(1)** a post-processing step to remove small islands or fill holes (from any classes), **(2)** the use of ensemble learning to increase robustness,[32] **(3)** the use of 3D convolutions to reduce uncertainties,[26] or **(4)** any recent deep learning tricks that have been proposed to improve segmentation accuracy.[33] In practice, this problem may never be eliminated; clinically, this could pose a major challenge, as early ODD signs could be labelled as segmentation errors. To better understand and address the extent of this phenomenon, it would be critical for us to reproduce this work in a considerably larger population with varied ODD severities, including both adults and children, considering all types (i.e., visible vs buried), and with varying OCT acquisition protocols.

The performance of the classification was excellent. AUCs of 0.99±0.01 were obtained for the detection of ODD (vs healthy+papilldema), 0.99±0.01 for the detection of papilledema (vs healthy+ODD), and 0.98±0.02 for the detection of healthy ONHs (vs ODD+papilledema). Our approach took full advantage of the Drusen score (typically high in ODD eyes, but zero or low in healthy and papilledema eyes), and of the prelamina swelling score (typically high in papilledema eyes, low in healthy eyes, and in-between in ODD eyes). We believe that our approach has the potential to establish OCT as the mainstay of diagnostic imaging for discriminating ODD from papilledema. As opposed to other techniques to detect ODD (e.g. B-scan ultrasonography, fundus autofluorescence),[34] OCT is the only technology that is able to provide 3D structural information about the ONH with high resolution regardless of ODD size, depth and degree of calcification.[5] Such information can readily be exploited by AI to ultimately provide the best diagnostic performance – but this would need to be carefully evaluated.

If OCT becomes the technology of choice for ODD and papilledema, our AI-based approach would also represent an advantage over pure 'human' interpretation of the OCT B-scans. This would not only benefit ophthalmologists, but also neurologists who might be less familiar with the rather complex structural organization of the ONH.[35] Many studies have advocated the merit of OCT to detect ODD regions and conglomerates,[4-6, 12-17] however to the best of our knowledge, none have used AI to perform a diagnosis while learning from the complex and deep anatomy of the ONH. If we are to solely rely on human (rather than machine) interpretation, multiple concerns arise. First, a clinician will need to manually investigate each B-scan individually (among up to several hundred depending on the OCT manufacturer); this is hard to achieve in busy clinics. Second, if the OCT acquisition protocol is not standardized, the chance of missing ODD regions and conglomerates is high in case of a large distance between B-scans. Third, for deeper ODD regions and conglomerates, where light attenuation is at its peak, there is a risk of not properly identifying ODD regions and conglomerates without the use of restoration tools such as compensation (as employed herein).[19] Fourth, clinicians will need to be trained to recognize complex ONH structures in OCT B-scans, and their expressions in the presence of machine-dependent artefacts; this is clearly a major endeavor. In the future, we aim to formally assess whether machine interpretation of OCT B-scans is in fact superior to that performed by humans.

Overall, we believe that our technology has the potential to reduce the number of false negatives, i.e. the papilledema cases that have been identified as ODD. Furthermore, we may be able to eliminate unnecessary, invasive, and expensive investigations in ODD patients. In the post COVID-19 era, it will become critical to minimize testing, to encourage telemedicine, and to develop novel home-monitoring solutions.[36] Our technology is fully aligned with these goals, and as OCT gets democratized worldwide, it may enter the home of the patients.

We also believe that our AI technologies could become invaluable for the pediatric population, where diagnostic uncertainty is even more common. Early in life, ODD are typically buried making them harder to detect clinically.[37] A quick non-contact mode of diagnosis like OCT would be preferred for this population,[38] and we have an opportunity to address this problem with our technology. However, it is yet unknown as to whether our approach would be directly transferrable to pediatric datasets, or whether re-training would need to be performed.

In this study, several limitations warrant further discussion. First, our sample size was relatively small, but it also needs to be emphasized that the prevalence of ODD is low (0.2% for ODD in an adult Chinese population;[39] 2% in an adult Danish population),[40] and that of papilledema much lower.[41] A small sample size limited us from performing a proper validation and test assessment on the classification problem, as is common in machine learning. Our AUCs may also have been inflated. it is important to keep in mind that the reported AUCs are only valid to this specific population and may not translate to other populations. In the future, it will be critical to validate our work in a much larger and heterogenous population (including all ages, ethnicities, and ODD types).

Second, ODD cases were confirmed by inspection of the OCT scans by clinician experts as per the consensus guidelines.[4] Other technologies, such as B-scan ultrasonography, or fundus autofluorescence imaging were not used to confirm the diagnosis.[7, 34]

Third, our segmentations excluded some potentially relevant tissues, including, but not limited to: retinal layers (other than RNFL, RPE, GCL and IPL), PHOMS,[31] the central retinal vessel trunk (and its branches),[30] and the border tissues of Elschnig and Jacoby.[42] A proper detection of such tissues may in turn improve the detection of ODD regions and conglomerates. This will be tested in the near future.

Fourth, our technology was only tested with one OCT device (Spectralis). While it was able to handle multiple scan sizes (all raster, but not diagonal), we have yet to make it device-agnostic, as we proposed in [26] for healthy and glaucoma eyes.

Fifth, we did not include cases of optic disc swelling which were unrelated to high intracranial pressure (i.e., occurring in ischemic, inflammatory, compressive anterior optic neuropathies), but they typically have different clinical profiles.[2] Also, we did not include cases of co-existing papilledema and ODD, which tend to be very rare. In the future, it will become important for us to develop a versatile tool that would handle multiple optic neuropathies.

Finally, we were unable to provide an additional validation of our segmentation approach by comparing our 'stained' OCT images to those obtained from histology. While this is would be extremely invaluable to understand what tissues are exactly being detected, in practice, it will remain a challenge to obtain human donor eyes exhibiting either ODD or previous signs of papilledema.

In conclusion, we propose herein novel OCT-based AI technology to discriminate ODD from papilledema with high accuracy. Our technology has strong potential to assist both ophthalmologists and neurologists to reduce misdiagnoses, and to establish OCT as the mainstay of diagnostic imaging for ONH disorders such as ODD in neuro-ophthalmology.

## Acknowledgment

We acknowledge funding from **(1)** the donors of the National Glaucoma Research, a program of the BrightFocus Foundation, for support of this research (G2021010S [MG]), **(2)** SingHealth Duke-NUS Academic Medicine Research Grant (SRDUKAMR21A6 [MG]) **(3)** Singapore National Medical Research Council (Clinician Scientist Individual Research grant CIRG18Nov-0013 [DM]), and **(4)** the Duke-NUS Medical School, Ophthalmology and Visual Sciences Academic Clinical Program grant (05/FY2019/P2/06-A60 [DM]).